\documentstyle[12pt]{article}
 
\begin{document}
\title{Discretizing $\upsilon$-Andromed{\ae}  planetary system}
\author{
A. G. Agnese  
\thanks{<agnese@genova.infn.it> INFN, Sezione di Genova} \and
R. Festa 
\thanks{<festa@fisica.unige.it> INFM, Unit\`a di Genova}
}
\date{\small Dipartimento di Fisica, Universit\`a di Genova, Italy}

\maketitle
\vspace{0.3cm}
\hrule
\begin{abstract}
Starting from a previously stated hypothesis concerning the discretization of the
orbits for periodic celestial motions, the mass of $\upsilon$ Andromed{\ae} and
the periods of its b- and d- companions are estimated using the accurately
known period of the massive
c-companion. Moreover, the periods and the orbital  major semiaxes 
of other hypotetically existing and not yet discovered smaller planets
of $\upsilon$ Andromed{\ae} are guessed.

\end{abstract}
\hrule
\vspace{0.5cm}
\section{Introduction}
In a previous work [1] the hypothesis was tested that periodic celestial 
motions comply with orbital discretization rules similar to those given by
Bohr-Sommerfeld for the Old Quantum Mechanics. In the case of a single planet 
gravitating in the field of a star of mass $M$ the discretization rules can be written
\begin{equation}
\label{rule}
\oint v_i\;dx_i = n_i \; 2 \pi \frac{G\;M}{\alpha_g c}\hspace{1cm} (i=1,2,3)\;,
\end{equation}
where $G$ is the gravitational constant, $c$ the light speed and $\alpha_g$
a dimensionless gravitational structure constant to be determined on the basis of the
observational data. From (\ref{rule}) the following
relations concerning the orbital major semiaxis $a_n$ and the corresponding
orbital periods $P_n$ can be easily deduced
\begin{equation}
\label{radiusperiod}
\left\{
\begin{array}{lcl}
a_n &=& n^2\; a_1\; ;\hspace{1cm}[a_1 = \frac{G M}{\alpha_g^2 c^2}]\\
P_n &=& n^3\;P_1\; ;\hspace{1cm}[P_1 = 2\pi\frac{GM}{\alpha_g^3 c^3}]\;.
\end{array}
\right.
\end{equation}
Applying this scheme to the solar sistem we obtained (see [1])
the following selection for the quantum numbers $n$:
Mercury (3), Venus (4), Earth (5),
Mars (6), Jupiter (11) , Saturn (15),
Uranus (21), Neptune (26), Pluto (30),
and the numerical specifications
\[
\left\{
\begin{array}{lcl}
\frac{1}{\alpha_g^\odot} &=& 2086 \pm14 \\
 & & \\
a_1^\odot  &=&  0.04297\;a. u.\\
 & & \\
P_1^\odot  &=& 3.269\;d\;,
\end{array}
\right.
\]
where the symbol $~^\odot$ refers to solar system related quantities.
The fit was performed both giving equal statistical weights to all planets 
and taking into
greater account Jupiter, which, owing to its mass, is
reasonably the less perturbed (almost isolated) planet.

Note that on account of Eqs. (\ref{radiusperiod}) 
for every celestial planetary system (of almost non interacting planets)
the equation holds
\begin{equation}
\label{periodmass}
\frac{P_n}{P_n^\odot} = \frac{M}{M^\odot},
\end{equation}
so that one can use the careful measurements referring to the solar system to perform
adequate comparisons.

On the basis of the same discretization pattern (\ref{rule}), the fundamental value for the dynamical spin
\[
J=\frac{GM^2}{2 \alpha_g^\odot c}
\]
was found, which surprisingly holds well for all celestial systems, from planets to
superclusters, on about twenty orders of mass magnitude, in agreement with
the Brosche-Wesson law ([2] [3] [4] [5]). Thus, on account of
this result, we provisionally
identified the  $\alpha_g^\odot$ with the gravitational structure constant $\alpha_g$.

\section{The extrasolar planets}
In a subsequent work ([6]), the previously stated discretization rules
were applied to the
recently discovered extrasolar planets (in fact about twenty star-planet
pairs). The comparison with experimental data gaves generally fair results,
mainly for planets with small orbital number ($n\le 2$). In fact, for most of such
systems, only the planet period was known with sufficient accuracy, 
whereas the star mass
was usually very roughly estimated from the star type:
of course, this fact produces some inaccuracy on the estimates,
mainly for large orbital number. When $n=1$ , one can compare the quantity
\begin{equation}
P_1^{(est)}=P_1^{\odot} \frac{M}{M^\odot}
\end{equation}
with the observed value $(P_1)^{(obs)}$.
In the following table we 
report some comparisons obtained using the previous equations and uptaded
observational values ([7]) :
\begin{center}
\begin{tabular}{||l||c|c|c|c|c|c|c||}
\hline\hline
Star       & Star type  &  P$^{(obs)}\;[d]$&$ a_1[a.u.]$ &P$^{(est)}\;[d]$\\
\hline
HD 187123  & G5         &  3.097           &0.042       & 3.19     \\
tau Bootis & F6IV       &  3.313           &0.046       & 3.50     \\
HD 75289   & G0V        &  3.508           &0.047       & 3.58     \\
51 Peg     & G2IVA      &  4.231           &0.051       & 3.88     \\
\hline
\end{tabular}
\end{center}
Note that even for some planet with $n=2$ (for instance,HD 192263,
see 
[8]) one would obtain a
fair agreement (P$^{(est)}=22.8\;[d]$  {\em vs} $P^{(obs)}=23.9\;[d]$ ) 
If we recall that the reported major semi-axes have been calcolated using
the (roughly estimated) star masses,
we can note that all the examined planets
satisfactorily follow our rule.
The only disagreement would occur  for the star-companion system
HD 217107 ($ (P_1)_{est}=5.32\;d$ {\em vs}  $P^{(obs)}=7.11\;[d]$ ).
\section{The $\upsilon$ Andromed{\ae} planetary system} 

In the last year, one of the previously discovered star-companion  pairs, namely 
$\upsilon$-Andromeda, has shown itself to be  a real planetary system, with three planets,
simply named b, c, and d (see [9]). Planet b, declared as 
very similar to Jupiter, is about $0.059\; a.u.$
from the star, and its period is $4.617\; d$. The mass of c-planet is about 
twice as the b-planet mass, and its orbital semiaxis 
and period are respectively
$0.83 \; a. u.$ and $241.2\; d$. The third planet, at a distance $2.5 \;a.u.$ from
$\upsilon$ Andromed\ae,  has a mass about four times that of the first and
a period is $ 1267\; d$ .

We recall that, for the solar system, a simple fit
based only on the almost unperturbed Jupiter (with $n=11$) 
gave reliable results for the whole solar system.
Likewise, we take the massive c-companion,
whose period ($P=241.2\;d$) is known with some accuracy, 
as the reference planet for 
the $\upsilon$ Andromed{\ae} system 
and consistently assign
$n=4$ to its orbit. Then, we can immediately calculate the $\upsilon$ Andromed{\ae} mass
$M^{\upsilon A} = 1.15\; M^\odot$ (Gray [10] estimated  
$M^{\upsilon A}= 1.20\;M^\odot$, whereas Ford et al. [11] 
declared a value $M^{\upsilon A}= 1.34\;M^\odot$).
Moreover, we simply get the period $P_1 = 3.76\;d$ for the b-companion and the
period $P_7 = 1289\;d$ for the d-companion, both values being highly consistent with
the observed values.

At this point, we can give a table with the guessed values of the periods and 
the major semiaxes of the $\upsilon$ Andromed{\ae} planetary system, including
the orbits of  hypotetically existing,
though not yet observed, smaller companions:

\begin{center}
\begin{tabular}{||l|c|c||}
\hline\hline
n & $P_n=n^3 P_1\; [d]$ & $a_n=n^2 a_1\; [a.u.]$ \\
\hline
1 (observed)   & 3.76     & 0.05 \\
2              & 30.1     & 0.20 \\
3              & 101      & 0.44 \\
4 (observed)   &  241     & 0.79 \\
5              &  470     & 1.23 \\
6              &  812     & 1.78 \\
7 (observed)   & 1289     & 2.42 \\
8              & 1985     & 2.75 \\
9              & 2711     & 3.46 \\
10             & 3760     & 4.30 \\
\hline\hline
\end{tabular}
\end{center}

\section{Conclusions}
Assuming that simple discretization rules (see [1]) hold for stable 
periodic celestial motions,
one is able  in particular
to calculate and compare the characteristic periods $P_n$ of newly discovered 
planetary systems of a star of mass $M$ using the relation
\begin{equation}
\label{Relation}
P_n = n^3 P_1^\odot \frac{M}{M^\odot}\;,
\end{equation}
where $P_1^\odot= 3.27\;d$ is a reference period related to the solar system,
$M^\odot$ is the solar mass, and $n$ an integer number.

However, the reasons why the mentioned discretization rules hold, very satisfactorily or 
almost
approximately, for a large group of periodic 
celestial motions are not at all clear at present. In our opinion, 
Nelson's stochastic mechanics 
([12], [13]) applied to the protoplanetary matter 
using the value 
$D = G M /2 \alpha_g c$ for the diffusion coefficient([1]),   
or fractal space-time schemes 
([14], [15])
are
today the best candidates to explain the effectiveness of our simple
discretization rules. Nevertheless, we suspect that equivalent results
could be obtained by looking for stable attracting orbits, in the
sense of the chaotic dynamics, for the strongly nonlinear gravitational motions.

We explicitely note that our mechanical scheme 
does not agree with
the recently proposed
theories of  the planet migration (see for instance [16]).
\vspace{1cm}

\noindent{\Large {\bf References}}

\vspace{0.4cm} 

\noindent [1] A.G. Agnese and R.~Festa.
Clues to discretization on the cosmic scale.
{\em Physics {L}etters {A}}, 227:165--171, 1997.

\vspace{0.3cm}
 
\noindent [2] P.~Brosche.
Zum {M}asse-{D}rehimpuls-{D}iagramm von {D}oppel- und
{E}inzelsternen.
{\em Astronomische Nachricten}, 286:241, 1962.

\vspace{0.3cm}

\noindent [3] P.~Brosche.
{\"U}ber das {M}asse-{D}rehimpuls-{D}iagramm von {S}piralnebeln und
anderen {O}bjecten.
{\em Zeitschrift f{\"u}r {A}strophysik}, 57:143, 1963.

\vspace{0.3cm}

\noindent [4] P.S. Wesson.
Self-{S}imilarity and the {A}ngular {M}omenta of {A}stronomical
{S}ystems: {A} {B}asic {R}ule in {A}stronomy.
{\em Astronom.Astrophys.}, 80:296--300, 1979.

\vspace{0.3cm}

\noindent [5] P.S. Wesson.
Clue to the unification of gravitation and particle physics.
{\em Phys. Rev. D}, 23:1730--1734, 1981.

\vspace{0.3cm}

\noindent [6] A.G. Agnese and R.~Festa.
Discretization on the cosmic scale inspired from the old quantum
mechanics.
{\em Adronic {J}ournal}, 8:237--254, 1998.

\vspace{0.3cm}

\noindent [7] J.~Schneider.
The {E}xtrasolar {P}lanets {E}ncyclopedia.
{\em INTERNET}, http://www.obspm.fr/encycl/catalog.html, 1999.

\vspace{0.3cm}

\noindent [8] N.~Santos, M.~Mayor, D.~Naef, F.~Pepe, D.~Queloz, 
S.~Udry, M.~Burnet, and Y.~Revaz.
A planet orbiting the star {HD} 192263.
{\em Cool {S}tars, {S}tellar {S}ystems and {S}un}, Tenerife, October
4-8, 1999.

\vspace{0.3cm}

\noindent [9] R.P. Butler, G.W. Marcy, D.A. Fisher, T.W. Brown, 
A.R. Contos, S.G. Korzennik,
P.~Nisenson, and R.W. Noyes.
Evidence for {M}ultiple {C}ompanions to {U}psilon {A}ndromedae.
{\em Submitted to {A}strophysical {J}ournal}, 1999.

\vspace{0.3cm}

\noindent [10] D.F. Gray.
{\em The {O}bservation and {A}nalysis of {S}tellar {P}hotospheres}.
{C}ambridge {U}niversity {P}ress, 1992.

\vspace{0.3cm}

\noindent [11] E.B. Ford, F.A. Rasio, and A.~Sills.
(preprint, {\em submitted to {A}strophysical {J}ournal}), 1998.

\vspace{0.3cm}

\noindent [12] E.~Nelson.
Derivation of the {S}chr{\"o}dinger {E}quation from {N}ewtonian
{M}echanics.
{\em Physical {Review}}, 150:1079--1085, 1966.

\vspace{0.3cm}

\noindent [13] Ph. Blanchard, Ph. Combe, and W.~Zheng.
{\em Matematical and {P}hysical {Aspects} of {S}tochastic
{M}echanics}.
Springer Verlag, Berlin, 1987.

\vspace{0.3cm}

\noindent [14] L.~Nottale, G.~Schumacher, and J.~Gay.
Scale relativity and quantization of the solar system.
{\em Astronomy and {A}strophysics}, 322:1018--1025, 1997.

\vspace{0.3cm}

\noindent [15] M.~Agop, H.~Matsuzawa, I.~Oprea, R.~Vlad, 
C.~Sandu, and C.Gh. Buzea.
Some {I}mplications of the {G}ravitomagnetic {F}ield in {F}ractal
{S}pace-{T}ime {T}heory.
{\em Australian J. of Physics}, 1999 (in press).

\vspace{0.3cm}

\noindent [16] N.~Murray, B.~Hansen, M.~Holman, and S.~Tremaine.
Migrating {P}lanets.
{\em Science}, 279:69--71, 1998.


\end{document}